 \documentclass[final,5p,times,twocolumn,sort&compress]{elsarticle}
 \journal{Nuclear Inst. and Methods in Physics Research, A}

\usepackage{amssymb}
\usepackage[british]{babel}
\usepackage{amsthm}
\usepackage{amsmath}
\usepackage{amsfonts}
\usepackage{graphicx}
\usepackage[table]{xcolor}
\usepackage{pdfpages}
\usepackage{multirow}
\usepackage{todonotes}
\usepackage{menukeys}
\usepackage{enumitem}
\setlist[enumerate]{itemsep=-6pt}
\usepackage[normalem]{ulem}
\usepackage{arydshln}
\usepackage{footnote}
\usepackage{textcomp}
\usepackage{hyperref}  
\hypersetup{colorlinks = true}
\setcitestyle{square}
\makesavenoteenv{tabular}
\makesavenoteenv{table}

\newcommand{\CDG}{\mathrm{CDG}}
\newcommand{\VGC}{\mathrm{VGC}}

\newcommand{\MFC}{\mathrm{MFC}}
\newcommand{\PS}{\mathrm{PS}}
\newcommand{\m}{\mathrm{m}}

\newcommand{\mm}{\mathrm{mm}}
\newcommand{\s}{\mathrm{s}}
\newcommand{\h}{\mathrm{h}}
\newcommand{\kg}{\mathrm{kg}}

\newcommand{\mbar}{\mathrm{mbar}}

\newcommand{\sccm}{\mathrm{sccm}}

\newcommand{\A}{\mathrm{A}}
\newcommand{\B}{\mathrm{B}}
\newcommand{\C}{\mathrm{C}}
\newcommand{\D}{\mathrm{D}}
\newcommand{\E}{\mathrm{E}}
\newcommand{\J}{\mathrm{J}}
\newcommand{\mol}{\mathrm{mol}}
\newcommand{\K}{\mathrm{K}}

\begin{document}
	\begin{frontmatter}
		
		
		\title{Design and Operation of a Windowless Gas Target \\Internal to a Solenoidal Magnet for \\ Use with a Megawatt Electron Beam}
		\author[add1]{S.~Lee\corref{cor1}}
		\ead{sangbaek@mit.edu}
		\author[add1]{R.~Corliss}
		\author[add1]{I.~Fri\v{s}\v{c}i\'{c}}
		\author[add5]{R.~Alarcon}
		\author[add7]{S.~Aulenbacher}
		\author[add1]{J.~Balewski\fnref{pre1}}
		\author[add3]{S.~Benson}
		\author[add1]{J.~C.~Bernauer\fnref{pre2}}
		\author[add2]{J.~Bessuille}
		\author[add3]{J.~Boyce}
		\author[add3]{J.~Coleman}
		\author[add3]{D.~Douglas}
		\author[add1]{C.~S.~Epstein}
		\author[add1]{P.~Fisher}
		\author[add3]{S.~Frierson}
		\author[add6]{M.~Gar\c con}
		\author[add3]{J.~Grames}
		\author[add1]{D.~Hasell}
		\author[add3]{C.~Hernandez-Garcia}
		\author[add2]{E.~Ihloff}
		\author[add1]{R.~Johnston}
		\author[add3]{K.~Jordan}
		\author[add3]{R.~Kazimi}
		\author[add2]{J.~Kelsey}
		\author[add4]{M.~Kohl}
		\author[add4]{A.~Liyanage}
		\author[add3]{M.~McCaughan}
		\author[add1]{R.~G.~Milner}
		\author[add1]{P.~Moran}
		\author[add4]{J.~Nazeer}
		\author[add1]{D.~Palumbo}
		\author[add3]{M.~Poelker}
		\author[add5]{G.~Randall}
		\author[add1]{S.~G.~Steadman}
		\author[add3]{C.~Tennant}
		\author[add2]{C.~Tschal\"{a}r}
		\author[add2]{C.~Vidal}
		\author[add1]{C.~Vogel}
		\author[add1]{Y.~Wang}
		\author[add3]{S.~Zhang}
		
		\fntext[pre1]{Present Address: NERSC Berkeley Lab, 1 Cyclotron Road, Berkeley, CA 94720-8150, USA}
		\fntext[pre2]{Present Address: Department of Physics and Astronomy, Stony Brook University, Stony Brook, NY 11794, USA
			and Riken BNL Research Center, Brookhaven National Laboratory, Upton, NY 11973, USA	}
		
		\cortext[cor1]{Corresponding author}
		\address[add1]{Laboratory for Nuclear Science, Massachusetts Institute of Technology, Cambridge, MA 02139, USA}
		\address[add2]{MIT Bates Research \& Engineering Center, Middleton, MA 01949, USA}
		\address[add3]{Thomas Jefferson National Accelerator Facility, Newport News, VA 23606, USA}
		\address[add4]{Department of Physics, Hampton University, Hampton, VA 23668, USA}
		\address[add5]{Department of Physics, Arizona State University, Tempe, AZ 85281, USA}
		\address[add6]{IRFU, CEA, Universit\'e Paris-Saclay, F-91191 Gif-sur-Yvette, France}
		\address[add7]{Institut f\"{u}r Kernphysik, Johannes Gutenberg-Universit\"{a}t Mainz, D-55128 Mainz, Germany}
		
		
		
		
		\begin{abstract}
			A windowless hydrogen gas target of nominal thickness $10^{19}$~cm$^{-2}$ is an essential component of the DarkLight experiment, which is designed to utilize the megawatt electron beam at an Energy Recovery Linac (ERL).  The design of such a target is challenging because the pressure drops by many orders of magnitude between the central, high-density section of the target and the surrounding beamline, resulting in laminar, transitional, and finally molecular flow regimes.  
			The target system was assembled and operated at Jefferson Lab's Low Energy Recirculator Facility (LERF) in 2016, and subsequently underwent several revisions and calibration tests at MIT Bates in 2017.  The system at dynamic equilibrium was simulated in COMSOL to provide a better understanding of its optimal operation at other working points.  We have determined that a windowless gas target with sufficiently high density for DarkLight's experimental needs is feasible in an ERL environment.
		\end{abstract}
		
		
	\end{frontmatter}
	
	
	\section{Introduction}
	\label{sec:introduction}
	The DarkLight\footnote{Detecting A Resonance Kinematically with eLectrons Incident on a Gaseous Hydrogen Target.} experiment has been proposed to search for a new, bosonic mediator $A^\prime$ between dark matter and the constituents of the visible matter in the universe, with mediator mass between 10 and 100~MeV, via direct production in electron-proton scattering \cite{Kahn2012,Balewski2013}.  A key requirement of the detector design is a multi-mbar, windowless, gaseous hydrogen target.  This target system was designed and developed at the MIT Bates Research \& Engineering Center and first operated at Jefferson Lab's (JLab) Low Energy Recirculator Facility (LERF) during a commissioning run in August 2016 \cite{Tennant:2018iyo}.  In the following year, it was reassembled at Bates to further test, calibrate, and improve the design.  This paper focuses on the technical description and the calibration of the DarkLight target system.
	
	\begin{figure}[!bh]
		\centering
		\includegraphics[width=90 mm]{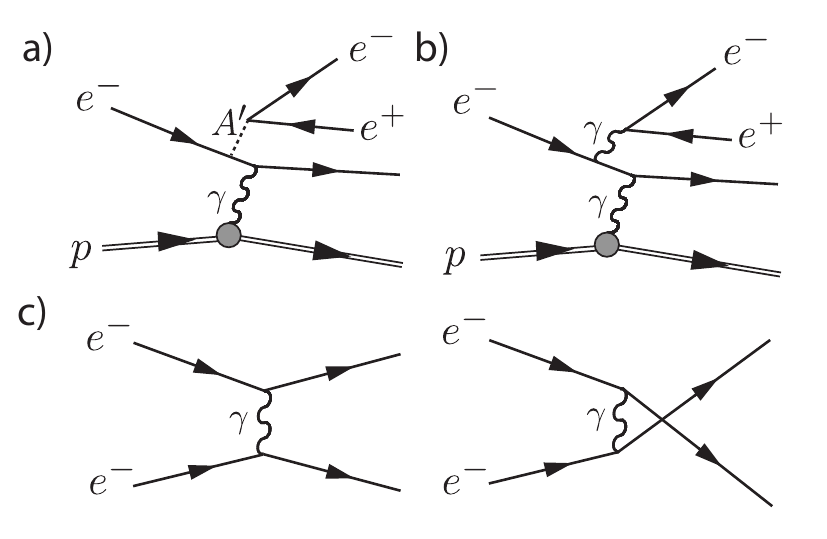}
		\caption{Feynman diagrams of the dominant reactions of (a) signal and (b) QED background.  Other signal and background reactions are illustrated at \cite{Freytsis:2009bh}.  The two diagrams on the bottom show (c) M\o ller scattering.}
		\label{fig:Feynman_diag}
	\end{figure}
	\begin{figure*}[!t]
		\centering
		\includegraphics[width=.95 \textwidth]{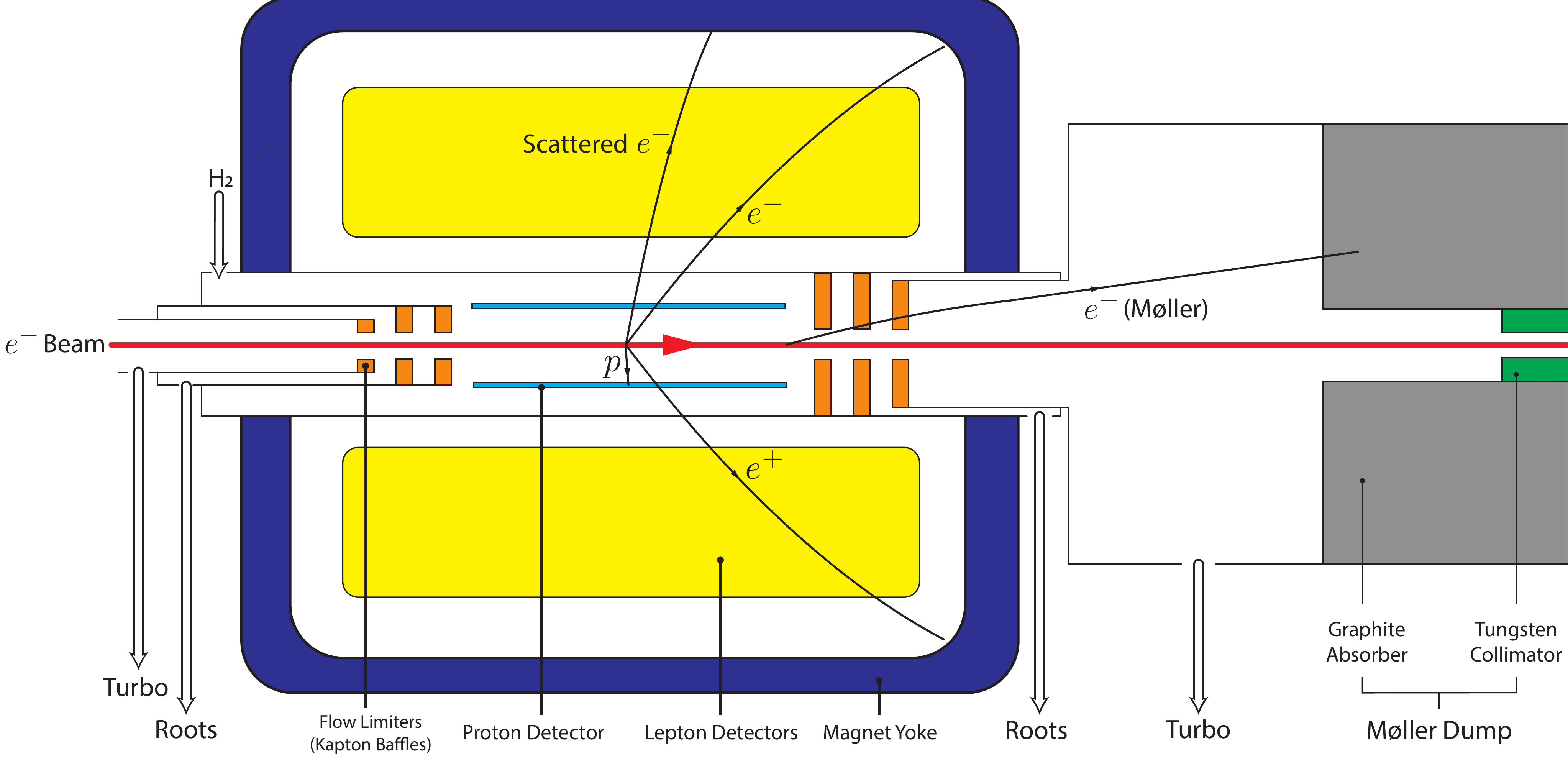}
		\caption{Schematic drawing of the overall DarkLight experiment (not to scale).  An ERL electron beam (red) enters from left and passes through the DarkLight target.  Lepton detectors (yellow) and a proton detector (light blue) are located within a solenoidal magnet (blue).  
			In the target system, gas flows from the hydrogen inlet (labeled `H$_{2}$', at left) through the central target region and out through a series of baffles (orange) to the up- and downstream beamline where it is removed by Roots pumps and turbopumps (captioned `Roots' and `Turbo').  
			Black lines show the trajectories of particles from a signal event (Fig.~\ref{fig:Feynman_diag}-a) at left, and a M\o ller scattering background event (Fig.~\ref{fig:Feynman_diag}-c) at right.  The forward electrons are captured by the M\o ller dump consisting of a graphite absorber (gray), and a tungsten collimator (green).  The downstream beam pipe must be wider than the upstream counterpart to prevent the M\o ller  electrons from hitting the pipe wall.}
		\label{fig:experiment_overview}
	\end{figure*}
	The search for non-gravitational evidence of dark matter has been an ongoing focus in particle physics, but no direct detection has yet been made.  Among the possible models, the dark sector model assumes a dark photon, $A'$, which interacts with both the standard model (SM) particles and dark matter.  A broad search has been made in the parameter space of the $A'$-SM interaction scale and the $A'$ mass by several experiments \cite{Lees2014, Agakishiev2014,Adare2015, Abrahamyan2011}.  The LHCb collaboration has also recently developed a scheme for an $A'$ search \cite{Ilten2018, PhysRevLett.116.251803}.  To date, much of the dark photon parameter space has been ruled out at 2$\sigma$; however, there are several low energy anomalies that hint at new physics beyond the standard model in these energy regions.  First, the anomalous magnetic moment of the muon has a discrepancy of 3.6$\sigma$ \cite{Bennett2006} with standard model predictions.  Additionally, a recent report claims a 6.8$\sigma$ anomaly in the decay of excited $^{8}$Be nuclei \cite{Krasznahorkay2016, Feng2016}, consistent with a new particle with mass near 17~MeV/c${}^2$.  Both of these anomalies could be explained by a more generalized version of a dark photon, where the couplings to different particle species are no longer identical.  Such a new {\it fifth force} would be evident in $e^+e^-$ decays in electroproduction experiments below pion threshold.  The DarkLight experiment is specifically designed to search for such decays in elastic electron-proton collisions with an incident electron energy of 100~MeV.  
	
	
	The signal of a new mediator boson produced in $ep$ scattering is an additional $e^+e^-$ pair in the final state (Fig.~\ref{fig:Feynman_diag}-a), with an invariant mass peaked sharply at the mass of the new particle.\footnote{This presumes there are no dark matter particles light enough to provide an alternate decay mode.}  The basic principle for detection is to track leptons in the final state with sufficient resolution to keep this peak narrow, and to accumulate enough statistics for the excess in a search window to exceed those expected by statistical fluctuations of the SM background (Fig.~\ref{fig:Feynman_diag}-b).
	The irreducible background comes from the SM process $e^-p \rightarrow e^-p \gamma^*\rightarrow e^-pe^+e^-$, but it is also critical to distinguish between true $e^-pe^+e^-$ final states and those faked by a random coincidence of elastic and M\o ller scattering (Fig.~\ref{fig:Feynman_diag}-c), both of which have rates much higher than the irreducible background.  Full reconstruction of the four-particle final state helps to discriminate against these backgrounds.

	In order to achieve the desired 1 $ab^{-1}$ integrated luminosity with a relatively low-density gas target (a few mbar), the experiment is designed to be operated at an Energy Recovery Linac (ERL) \cite{Kahn2012, Balewski:2014pxa}.  Such an accelerator provides an energy-efficient, high-current beam that is tolerant of higher areal density targets than traditional storage rings, and significantly higher beam currents than available in non-recirculating beams.  The high beam power at an ERL provides an additional constraint on detector design\,---\,any material interacting with the beam or beam halo may see significant heating and scatter electrons at rates that overwhelm detector elements.

	This article continues as follows:  Section \ref{sec:Overview of Detector Concept} provides an overview of the DarkLight detector, while Section \ref{sec:Design of DarkLight Hydrogen Gas Target} gives greater detail on the design of the target system itself.  Section \ref{sec:Operation at JLab} discusses the outcome of the commissioning run at JLab in 2016, and Section \ref{sec:Target Calibration at Bates} describes the calibration of the apparatus after reassembling at Bates in 2017.  The appendix lays out the details of the calibration tests.
	
	\section{Overview of the Detector Concept}
	\label{sec:Overview of Detector Concept}
	
	In order to demonstrate that it was a suitable beam for precision physics, the LERF was first tested to prove that it was sufficiently stable.  Stable beam operation through a millimeter scale aperture was demonstrated at the facility (then called the Free Electron Laser) in 2012, with a 5~mA, 100~MeV, 430~kW continuous wave beam \cite{Alarcon2013-2,Alarcon2013,Tschalar2013}.  Further test of beam stability and energy recovery in the presence of DarkLight major components, as shown in Fig.~\ref{fig:experiment_overview}, was done during the commissioning run in 2016 (see Section \ref{sec:Operation at JLab}).
	
	The DarkLight design consists of a 0.5 Tesla solenoidal magnet, which houses the hydrogen gas target, a cylindrical proton detector, and several layers of cylindrical lepton detectors for particle tracking.\footnote{Developing a lepton tracker of helical-shape is proposed at \cite{Wang:2018tkz} for a future experiment.}  All of these elements are designed to minimize material exposed to the beam, and to minimize and control reducible backgrounds as much as possible.
	
	The solenoidal magnetic field that provides analyzing power for particle momentum is matched to the diameter of the target chamber, so that electrons from M\o ller scattering are kinematically incapable of striking the outer walls of the chamber.\footnote{The possible effects of radiative corrections on this constraint motivated
		an independent measurement performed at MIT's High Voltage Research Lab \cite{CharlesThesis}.}  This results in a ``M\o ller cone'' of forward-going electrons that is absorbed by a downstream M\o ller dump.
	
	Electron elastic scattering cannot be eliminated in this fashion, since these electrons carry close to the full beam energy regardless of scattering angle.  While scattering from the active regions of the target cannot be reduced, scattering from even a thin target window would be a substantial additional rate.  Further, the destructive heating of any window via the power deposition of the megawatt electron beam makes a windowless target chamber mandatory.
	
	\subsection{M\o ller Dump}
	\begin{figure}[!h]
		\centering
		\includegraphics[width=90 mm]{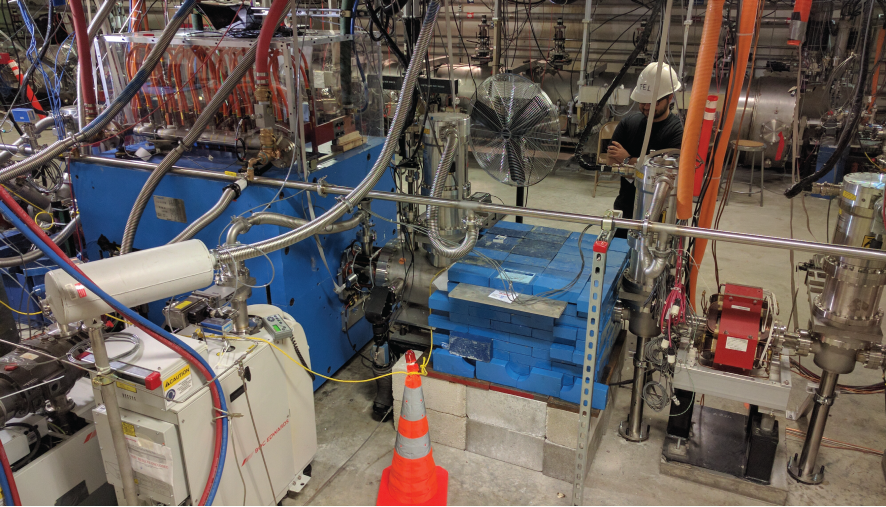}
		\caption{Photo of the DarkLight experiment installed in the LERF in August 2016.  The beam traverses from the left through the large blue solenoidal magnet.  In the center right is the M\o ller dump.  The carbon cylinder itself is located inside additional lead shielding (stacked blue bricks), downstream of the magnet yoke (large blue cube).  The Roots pump, used to differentially pump between target baffles, is in the foreground at left.}
		\label{fig:Moller dump}
	\end{figure}
	
	Though the solenoid controls the M\o ller envelope within the target volume, these trajectories expand once they leave the magnet.  An absorber that covers that solid angle is necessary to manage radiation levels in the beam hall.  
	The M{\o}ller dump consists of a 50~cm long graphite
	cylindrical shell with an inner diameter of 3~cm and outer diameter of  30~cm (Figs.~\ref{fig:experiment_overview} and \ref{fig:Moller dump}).  A tungsten collimator (length 15~cm, inner diameter 1~cm) is placed within the inner diameter of the graphite at the downstream end.  The low-$Z$ outer shell has a smaller $dE/dx$ but a smaller cross section for hard scattering, which stops electrons more gradually, and also helps attenuate those backscattered from the inner collimator.  Geant4 simulations of the design show that though it reduces the forward flux of electrons by several orders of magnitude, it produces significant secondary gammas, some small fraction of which will scatter back into the detector.  For this reason, the M\o ller dump is placed as far downstream as practical, balancing the anticipated photon rate against the clearance needed for other instruments along the beamline.  
	

	\section{Design of the DarkLight Target System}
	\label{sec:Design of DarkLight Hydrogen Gas Target}
	
	The windowless gas target was designed to maximize its density, while allowing the detection of low-energy ($\sim$1 MeV) recoil protons  and simultaneously minimizing the rate from beam-gas interactions outside of the target region itself.  Hydrogen was chosen as the target material to maximize the recoil kinetic energy.  
	
	The chamber itself is an aluminum tube with an inner diameter of 146~mm and thickness of 2 mm.  A thinner, beryllium tube was proposed for the future, but the original prototype used aluminum for cost and simplicity.
	The nominal areal thickness of the target is 10$^{19}$~cm$^{-2}$, corresponding to a 60~cm target at approximately 3.3~mbar.  Beamline constraints require the pressure outside the target to be at the 10$^{-7}$~mbar level so that the pressure near the cryomodules remains less than 10$^{-8}$ mbar.  Therefore, most of the hydrogen flow must be captured by pumps immediately up- and downstream of the experiment.  This requires a series of pumps at the highest possible effective pumping speeds combined with the lowest possible conductance to the beamline.
	
	To minimize conductance along the beamline, a flow-limiter is needed on either side of the target chamber.  A thin straw would minimize conductance, but has a significant radiation thickness for electrons emerging from the beam at small angles.  In simulations, this design was shown to dramatically increase the rate of multiply-scattered M\o ller and beam halo electrons striking the detectors and the M\o ller dump.  Instead, the chamber uses a set of Kapton baffles with small inner apertures and large outer diameters.  These have a larger conductance, but present a minimal amount of material for small-angle scattering.
	\begin{figure}[!h]
		\centering
		\includegraphics[width=90 mm]{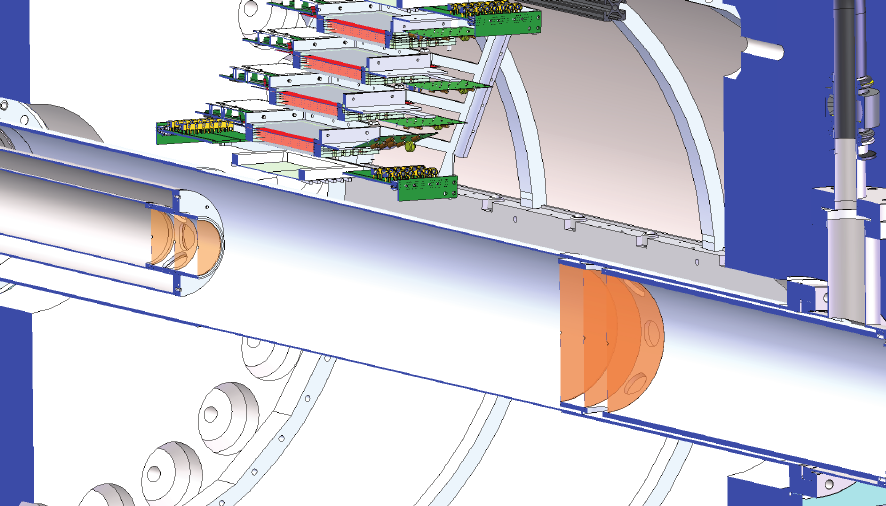}
		\caption{Target chamber with three Kapton baffles
			at its ends.  The two innermost baffles are 60~cm apart.  With a pressure of 3.3~mbar, this region corresponds to an areal thickness of 10$^{19}$~cm$^{-2}$.}
		\label{fig:baff}
	\end{figure}
	
	The 60~cm long target volume is enclosed by a set of baffles on each end.  Each of these sets consists of three 130~$\mu$m thick Kapton sheets spaced 38.1~mm apart.  Each sheet has a 3~mm diameter aperture, all of which are mutually aligned to allow the passage of the beam  (see Fig.~\ref{fig:baff}).  In order to increase the pressure gradient, the region between middle and outermost baffle is differentially pumped via a double-walled cylinder that connects it to a Roots-type blower (RTS).\footnote{Every RTS mentioned in this paper is Edwards iH1000 whose pumping speed depends on inlet pressure (see Section \ref{subsec:Pumping Speed of the Roots Blower iH1000}).}
	
	\begin{figure*}[!ht]
		\centering
		\includegraphics[width=.95 \textwidth]{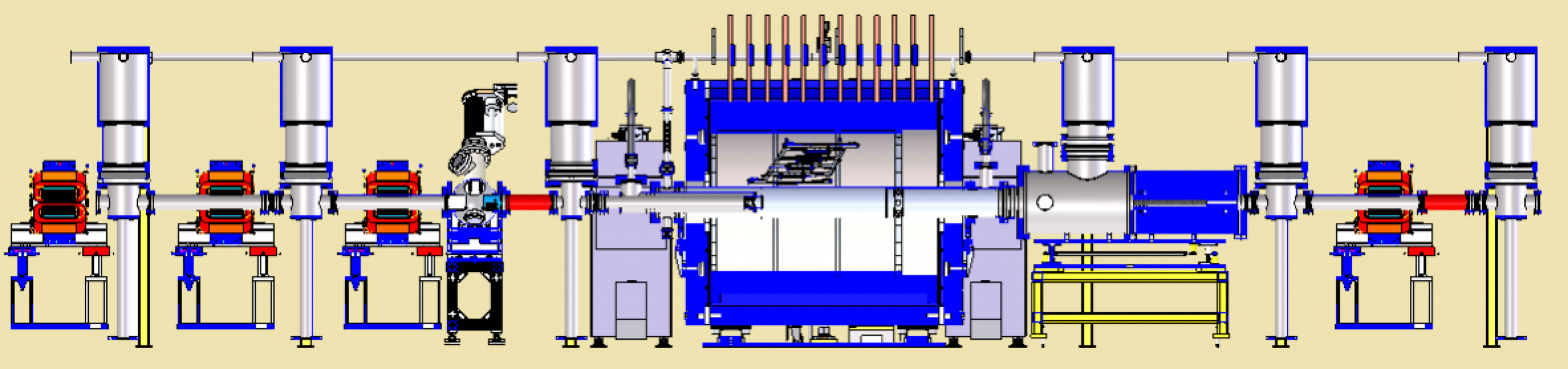}
		\caption{Layout of the DarkLight windowless gas target 
			system during the commissioning run at LERF.}
		\label{fig:xs-LERF}
	\end{figure*}	
	To maximize effective pumping speed, the annular region should have as large a conductance as possible.  On the downstream side, however, the annular region of the double walled pipe is narrower compared to the upstream side, to maintain safe clearance between the M\o ller envelope and the beam pipe.  For the commissioning run in 2016, the upstream annular cylinder had a length of 63.8~cm, an inner diameter of 63.5~mm, and an outer diameter of 97.4~mm, while the downstream cylinder was 48.8~cm long, with inner diameter 142.2~mm and outer diameter of 146.1~mm.  
	Beyond the baffles in both directions, the remaining gas flows into the beampipe, which is pumped on by a series of 3 Osaka TG1100M turbomolecular pumps.  These turbopumps operate at a constant pumping speed of about 3,000~$\m^3/\h$ for hydrogen at their operating pressure between about 10$^{-3}$~mbar and 10$^{-9}$~mbar \cite{TG1100M}.
	All six turbopumps are backed by a single Edwards QDP40 roughing pump.

	The gas in the target system is expected to go through multiple flow regimes due to the pressure change between the inlet and the beamline.  Hence, predicting the absolute pressures in the system using heuristic models is not necessarily reliable, and there is a need for prototypes and simulations.  
	A finite element analysis (FEA) tool can help to map the gas pressure profile by matching physical conditions from the sensors.  For DarkLight, the COMSOL Multiphysics software was taken as an FEA tool, allowing one to extrapolate to other working points.  Properly vetted, this tool allows some optimization of further proposed changes to the target system before implementing those changes in a physical prototype.
	
	\section{Operation at JLab}
	\label{sec:Operation at JLab}

	The system described in the previous section (shown in Fig.~\ref{fig:xs-LERF}) was installed at the LERF in 2016 \cite{Tennant:2018iyo}.  This included the target chamber mounted inside a solenoid magnet and accompanied by a prototype detector telescope \cite{Liyanage:2018wup}.  Two Edwards Roots pumps providing differential pumping were located alongside the magnet yoke, while the six Osaka Turbopumps\footnote{Repurposed from the OLYMPUS experiment} were mounted vertically on pump stands installed up- and downstream of the experimental area.
	
	During operation, hydrogen gas was fed into the target chamber and flowed out through the baffle apertures.  The flow of the hydrogen gas into the target chamber was controlled and monitored via a mass flow controller (MFC).  The pressure of the hydrogen gas inside the chamber, as well as at numerous points in the system,  was monitored with 16 vacuum gauges: 3 capacitance diaphragm gauges (CDG), 3 mini-convectrons and 10 Bayard-Alpert Pirani gauges (BPG).  The vacuum gauges were also part of a hardware interlock system to automatically protect the other sections of the accelerator and the turbopumps from unexpected over-pressuring.
	
	\begin{figure}[!h]
		\centering
		\includegraphics[width=90 mm]{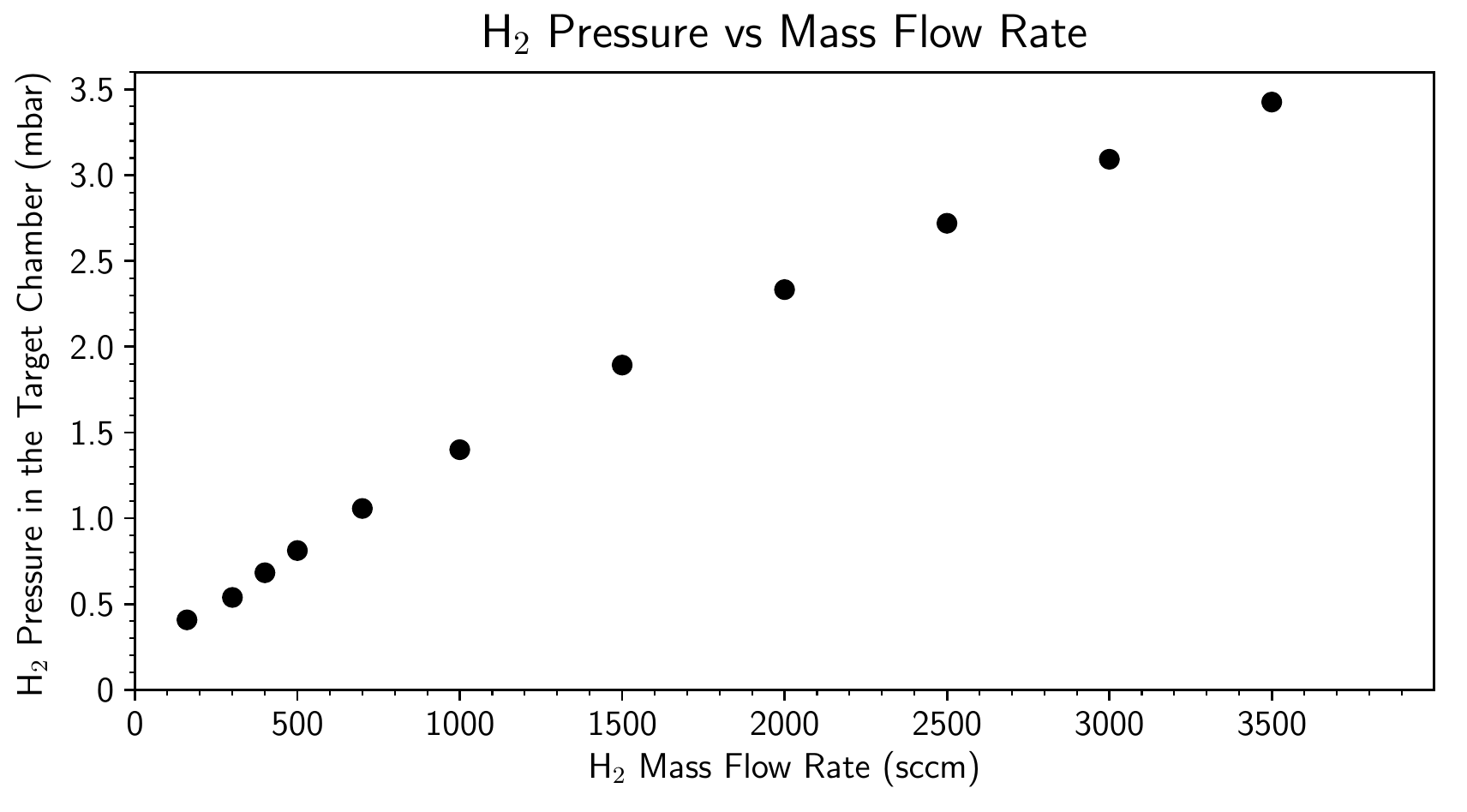}
		\caption{Hydrogen gas pressure as a function of the 
			mass flow.}
		\label{fig:bar1}
	\end{figure}
	
	Once assembled, we performed a series of tests of the gas system.  Due to the limited beam time available, several issues with the downstream system could not be repaired.  Damage to one of the downstream baffles likely increased conductance, and one of the three downstream turbopumps failed and had to be valved off.  
	First, we tested the maximum tolerable pressure in the system.  Valved off from the accelerator, the flow into the target was increased in steps and the pressure allowed to equilibrate.  Fig.~\ref{fig:bar1} shows the gas pressure in the target chamber as a function of the hydrogen flow.  The maximum tested 
	gas flow was 3500~sccm, at which the pressure at the first working downstream turbopump reached almost 0.13~mbar, posing a risk of damage to the pump.  For this value of the flow, we measured a hydrogen pressure of 3.43~mbar in the target chamber.  

	The system was then operated at a gas flow of 950~sccm for 1 hour and 15 minutes to test the long term stability.  The pressure in the target chamber remained stable at 1.35~mbar without any significant fluctuation.  
	
	The last test was to explore the stability of the beam in the presence of the target, and was performed with the rest of the accelerator exposed to the DarkLight gas target system.  This also allowed us to record the pressure effects along the beamline.  Preliminary tests showed an increased pressure  farther up and downstream of the experimental region, so these beam-on tests were operated at a very low gas flow setting, corresponding to a pressure of 0.408~mbar in the target chamber.  The limiting factor here was the pressure near the upstream ion pumps, which were used to maintain the high vacuum inside the accelerator beampipe.  The ion pumps closest to the target system were turned off, since the pressure at their position was above 10$^{-6}$~mbar while other pumps further away were kept on.  During several hours of run in this mode, slow buildup of pressure was observed at their positions.\footnote{During a longer, data-taking run, these pumps would be removed from the beamline.}
	
	Table \ref{labelT1} shows the calculated thickness of the hydrogen target for the above-described modes of operation and also gives the thickness of the gas targets used in the PRad \cite{PRad} and OLYMPUS \cite{Olympus} experiments.  Our lowest gas thickness is comparable with that used in the PRad experiment, and almost 3 orders of magnitude larger than in the OLYMPUS experiment.  
	
	In addition to the tests of the gas system, the commissioning run also demonstrated the LERF could be operated in energy-recovery mode in the presence of the energized solenoid (performed before the installation of the gas system), and provided the opportunity to test detector technologies intended for the full experiment.  Unfortunately, it was not possible to correct a misalignment of the baffles during the run period, so a full test of energy recovery in the presence of both gas and magnetic field was not possible.  Despite this, many possible improvements to the system were identified that would allow the DarkLight internal target to be operated at higher densities.
	
	\begin{table}[h]
		\caption{Thickness of the DarkLight gas target at pressures of 0.408, 1.35 and 3.43~mbar at room temperature, assuming an effective target length of 60~cm with comparison to the target thickness in the PRad \cite{PRad} and the OLYMPUS \cite{Olympus} experiments.} 
		\begin {center}
		\begin {tabular} {l|c|c|c}
		\hline
		\multirow{ 2}{*}{Experiment} & Pressure (mbar)/   &  Length & Thickness  \\
		&   Temperature (K)  & (cm)        & (cm$^{-2}$) \\
		\hline
		\multirow{ 3}{*}{DarkLight} & 3.43/ 293.15 & \multirow{ 3}{*}{60}   & $1.0 \times 10^{19}$   \\
		& 1.35/ 293.15 &  & $4.2 \times 10^{18}$ \\
		& 0.408/ 293.15 & & $1.2 \times 10^{18}$ \\
		\hline 
		PRad &  0.43/ 25 & 4 &  $9.9 \times 10^{17}$   \\
		\hline
		OLYMPUS &  $2.6\times10^{-4}$/ 75 & 60 &  $3.1 \times 10^{15}$   \\
		\hline
		\end {tabular} 
		\end {center}
		\label{labelT1}
		\end {table}
		
		\section{Target Calibration at Bates}
		\label{sec:Target Calibration at Bates}
		\subsection{Experimental Layout}
		\label{subsec:Experimental Layout}
		
		\begin{figure*}[ht!]
			\centering
			\includegraphics[width=.95 \textwidth]{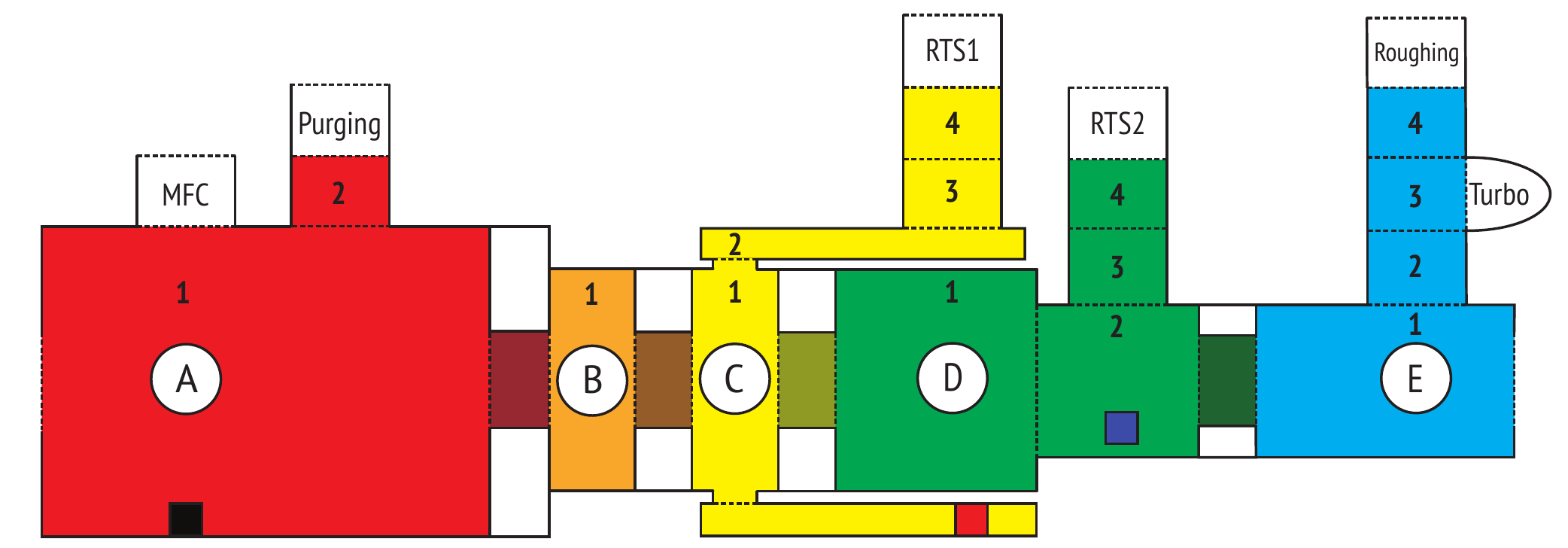}
			\caption{Schematic diagram of target system (not to scale).  Each region that hydrogen can populate is marked with distinct colors and surrounded by dotted border lines.  The baffles are shaded with a slightly darker color and surrounded by white rectangles.  Other boxed white annotations show the location of various external connections.  The black, red, and blue rectangles show the approximate locations of CDG 1, 2, and 3.  The color expression will be kept the same in this paper.
			}
			\label{fig:cartoonish}
		\end{figure*}
		
		
		The downstream portion of the target chamber was reassembled in 2017 at Bates in order to test repairs and modifications and conduct a series of calibration tests.  During reassembly, the damaged baffle was replaced and the inner pipe was exchanged for a narrower-diameter pipe to improve conductance of the annular region to the existing Roots blower (RTS1).  The inner diameter of this region decreased from 142.2~mm to 126.2~mm, and the openings connecting it to the intra-baffle region were enlarged.  Outside the target chamber and existing baffles, another Roots blower (RTS2 in Fig.~\ref{fig:cartoonish}) was added to the system, and an additional baffle with 3~mm aperture was newly installed between the new Roots and the turbomolecular pumps.   The M\o ller dump and remaining two turbopumps were integrated into the LERF beamline and so were not part of this assembly.
		
		A cartoon of the portion of the system reconstituted at Bates, showing the locations of the mass flow controller, pressure gauges, pumps, and baffles, is shown in Fig.~\ref{fig:cartoonish}.  CDGs were used to monitor the pressure in regions A, C, and D, and BPGs monitored regions D and E.  The nominal pressure uncertainty for each CDG is $\sigma_{P}=0.22~\% \times P_{meas.}^{\CDG}~\oplus$ 6.94$\times$10$^{-3}$~mbar \cite{CDG025D, VGC503}.  The mass flow uncertainty of the MFC is $\sigma_{\dot{m}}\sim5.32~\% \times \dot{m}_{meas.} \oplus 10 ~\sccm$ \cite{246C, GE50A}.

		\subsection{Overview of Tests}
		\label{subsec:Overview of Tests}
		In order to validate the pressure profile of the rebuilt system, we performed a series of tests that verified the calibration and performance of the pressure gauges, mass flow controller, and pumps.  
		These consisted of the following tests:
		\begin{enumerate}
			\item Pressure gauges were checked against one another at static pressure (Section~\ref{subsec:Test 1: Static Pressure Tests}).
			\item Calibration of the MFC was checked (Section 
			\ref{subsec:Test 2: Verifying MFC Calibration})
			\item Pumping speeds of the Roots pumps were checked in dynamic equilibrium (Section \ref{subsec:Test 3: Equilibrium Pumping Tests})
		\end{enumerate}
		
		For the first test, the calibration of each pressure gauge is checked against the others by filling the chamber to several different pressures (roughly 1.3, 4.0, and 6.7~mbar) and allowing it to equilibrate there.  Each step of this test is followed by a step of the second test, in which the chamber is pumped down to the 10$^{-3}$ mbar pressure level, after which the valves to the Roots and backing pumps are closed, and all pumps are turned off.\footnote{There is no valve between the turbopump and the target chamber, so the internal volume of that pump must also be considered in these tests.  The pump is, of course, off.} The MFC is opened to a fixed value, and the pressure at each gauge is recorded as a function of time.  
		
		For the third test, a subset of the pumps is on and the MFC is opened to various flow rates long enough for the system to settle into a steady-state, i.e.\ dynamic equilibrium.  The pressure at each gauge is recorded as a function of the mass flow setting.  We perform this test independently of the first two.
		
		A more thorough description of these tests and analysis of their results is provided in the following subsections.
		
		\subsection{Calculation of Internal Volume}
		\label{subsec:Calculation of Internal Volume}
		
		\begin{table}[h]
			\caption{Volumes of five regions pertinent to these tests A, B, C, D, and E. We assume conservative uncertainties arising from deviations from the nominal dimensions and so assign an overall uncertainty of 5~\%.} 
			\begin {center}
			\begin {tabular} {c|c}
			Component & Volume ($10^{-3}~\m^{-3}$)\\
			\hline
			A&19.5\\
			B&0.44\\
			C&3.2\\
			D&8.4\\
			E&12.3\\
			\hline
			Total&44.4\\
			\hline
			\end {tabular} 
			\end {center}
			\label{labelT2}
			\end {table}
			To check the calibration of the MFC, the total volume of the system must be known (Table \ref{labelT2}). Though dominated by the main body of the target (region A1 of Fig.~\ref{fig:cartoonish}), there are many additional regions that need to be correctly treated:
			
			\noindent 1.  The volume of the hydrogen gas feed line is so small that it is not included in the volume estimation, but the purging pipe is significantly larger.  The 1~m long and 2.54~cm wide pipe is valved at the far end, just before the purging pump, and so its volume must be included in calculations.
			
			\noindent 2.  Most components in Fig.~\ref{fig:cartoonish} can be modeled with concentric cylinders, but the angle valves (located at C4 and D4 of Fig.~\ref{fig:cartoonish}) involve cylinders intersecting at right angles.  The volumes of these elements in their open and closed states were estimated using numerical integration.
			
			\noindent 3.  There is no valve between the target chamber and the turbopump (located at E3).  The pertinent valve is instead on the pipe (E4) connecting the turbopump to its backing pump.  The common turbomolecular pump consists of multiple rotor blades and stator blades inside a vacuum chamber.  The empty space in this chamber is the second largest contribution to the total volume of the target system, and is approximately $6\times10^{-3}~\m^3$ (6 liters) \cite{privatecom}.  We assign a conservative uncertainty to this volume of 5~$\%$.
			
			\subsection{Static Pressure Tests}
			\label{subsec:Test 1: Static Pressure Tests}
			In between other tests, we allowed the chamber to settle at various pressures nine times, three times each at around 1.3, 4.0, and 6.7~mbar, from which we verify the consistency of the pressure measurements from each gauge. Along with uncertainties calculated from the manufacturer's manual \cite{CDG025D, VGC503},
			\begin{align}
				\label{eq:mean and uncertainty of pressure C}
				P_\C/P_\A =& 0.9976\pm0.0044\\
				\label{eq:mean and uncertainty of pressure D}
				P_\D/P_\A =& 0.9982\pm0.0044.
			\end{align}

			\subsection{Verifying Mass Flow Controller Calibration}
			\label{subsec:Test 2: Verifying MFC Calibration}
			Safety limits at the LERF required mixing the exhaust hydrogen from the pumps with nitrogen with a maximal fraction of 4~\% hydrogen in the mixed exhaust.  In addition to providing a stable flow that maintains a stable pressure in the system, the MFC allows us to explicitly ensure this safety condition.  It also allows us to compare data from the prototype with heuristic models and simulations.  In this section, we introduce how to verify the calibration of the MFC relying only on pressure measurements and hence without assuming a particular performance of the pumps.  This is conceptually straightforward: the mass flow into a closed system of fixed volume is related to the rate of change of the density of the gas.  Hydrogen gas below 10~mbar has a viscosity low enough to be treated as an ideal gas, allowing us to connect this to the pressure and temperature of the gas.  We take the temperature $T$ of the flowing gas to be 15~$\pm~5~{}^{\circ}$C.  The gas is at low pressure, and, even if there is adiabatic cooling at the inlet, will reach thermal equilibrium with the chamber after a few collisions with the wall.  Since the chamber does not accumulate frost, the temperature cannot be below about 280~K.
			
			Measurement of the pressure as a function of time allows us to derive the estimated mass flow rate,
			$\dot{m}_{\mathrm{Est}}$:
			\begin{align}
				\label{eq:massfloweqnsimple}
				\dot{m}_{\mathrm{Est}}&=\frac{d(\rho(T) V)}{dt}
				= V\frac{T_0}{T}\frac{d(\rho_0 P/P_0)}{dt}
				=V\frac{T_0}{T}\frac{\rho_0}{P_0}\frac{dP}{dt},
			\end{align}
			where $\rho$ is the mass density of hydrogen gas of pressure $P$, and $\rho_0=0.089~\kg/\m^3$ at standard temperature and pressure, $T_0=273.15$~K and $P_0=1000$~mbar.
			For a system that has multiple regions with different pressures, we sum the contributions for each region:
			\begin{equation}
				\label{eq:massfloweqn}
				\dot{m}_{\mathrm{Est}}=\frac{T_0}{T}\frac{\rho_{0}}{P_{0}}\sum\limits_{i\in \{\A,\B,\C,\D,\E\}}V_i\frac{dP_{i}}{dt} ,
			\end{equation}
			where $V_i$ and $P_i$ are the volume and pressure of each region.
			
\begin{figure}[!h]
				\centering
				\includegraphics[width=90 mm]{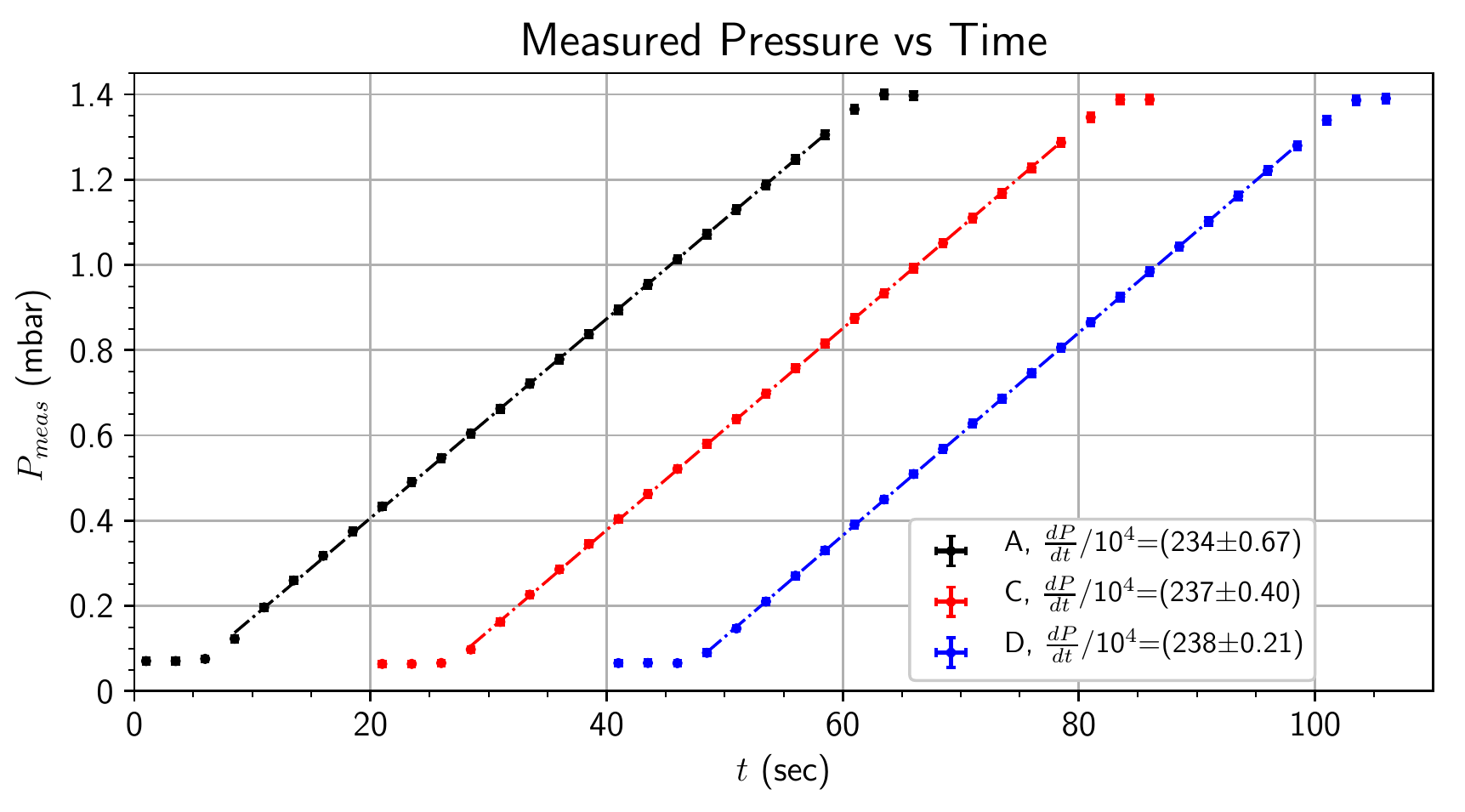}
				\caption{A representative data set of pressure as a function of time for the MFC tests.  Nonlinear regions at the beginning and end of the time series correspond to the MFC opening and closing.  Subscripts refer to the different pressure gauges.  For clearer visualization, the data are shifted by 20 sec per gauge.}
				\label{fig:TimeandPressure}
			\end{figure}

			The chamber was initially pumped down to $\sim$1~mbar via one of the Roots pumps, after which valves C4 and D4 were closed and the MFC was opened to a fixed setting.  This procedure has a ramp and settling time before the flow stabilizes at the desired value, which is quite short compared to the duration of the test.  The effect can be seen in Fig.~\ref{fig:TimeandPressure}.  As the chamber filled, the pressures in regions A, C, and D were recorded directly, while values in B and E must be inferred from measurements of the adjacent regions: $P_{\B}$ is taken to be the average of $P_{\A}$ and $P_{\C}$; the result is not particularly sensitive to this choice, since $V_{\B}$ is around 1~\% of the entire volume.  The value of $P_{\E}$ will depend on the test performed.  If the turbopump is not active, it will be the same as $P_\D$ in the static limit.  The estimated mass flow in sccm, as inferred from the temperature (in K), calculated volumes (in m$^3$), and measured pressures (in mbar) in the system, can be expressed as follows:
			
			\begin{align}
				\label{eq:massfloweqnAtest}
				\dot{m}_{\mathrm{Est}} = 1.62\times10^{7}~T^{-1}&\left(  V_{\A}\frac{dP_{\A}}{dt}+V_{\B}\frac{1}{2}\left(\frac{dP_{\A}}{dt}+\frac{dP_{\C}}{dt}\right) \right.  \nonumber \\
				&~ \left.  {}+V_{\C}\frac{dP_{\C}}{dt}+(V_{\D}+V_{\E})\frac{dP_{\D}}{dt}\vphantom{\frac12}\right).\end{align}
			
			The pressure time series measurements were fit to a linear pressure model, $P_i(t)=A_it+B_i$.  In addition to the pressure uncertainties (described in Section \ref{subsec:Experimental Layout}), a timing uncertainty arises, conservatively set at 1/60~s, from possible aliasing effects of the 30~Hz data recording.  An overall time delay in the pressure readings would only shift the entire curve, and hence will not impact the fit results.  The contribution of timing resolution to fitting is small compared to the uncertainty in the free space of the turbopump, where 5~$\%$ of the total internal volume of the pump corresponds to 1.8~$\%$ of the entire target system volume.
			\begin{figure}[!h]
				\centering
				\includegraphics[width=90 mm]{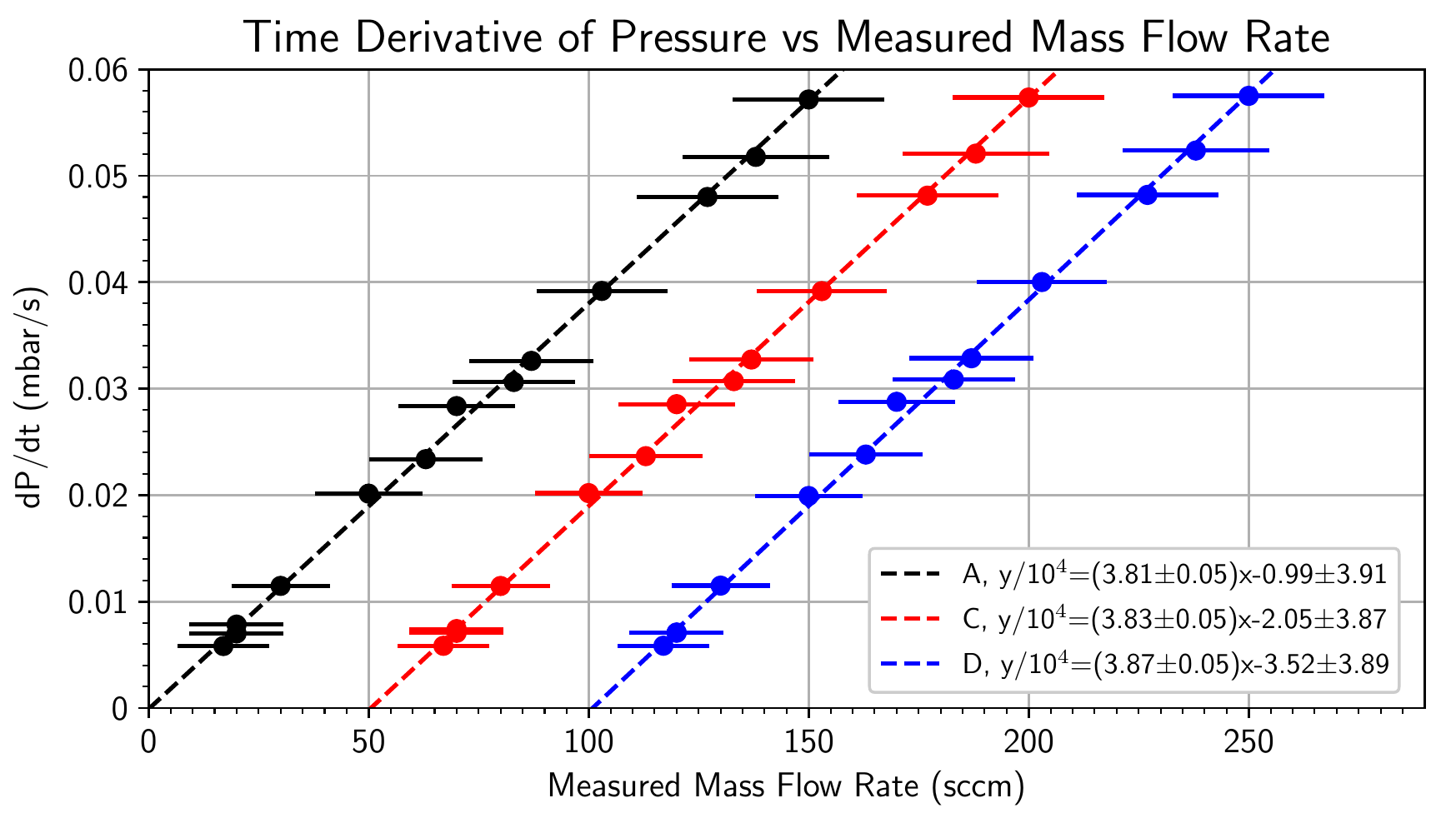}
				\caption{$dP/dt$ measured by gauges in regions A, C, and D as a function of MFC readout.  Error bars show the systematic uncertainty of the readout.  The MFC data are shifted by 50~sccm per gauge for clarity.}
				\label{fig:RawdPdt}
			\end{figure}
			
			The pressure was ramped 15 times at different MFC settings in order to check the linearity of the MFC calibration.  All of these produced linear plots, an example of which is shown in Fig.~\ref{fig:TimeandPressure}.  Collecting the coefficients from each of these fits, we confirm that the time rate of change of the pressure as measured by each gauge is linear with the dial setting of the MFC (Fig.~\ref{fig:RawdPdt}).  Data from different gauges are shown with different colors and different abscissa offsets.
			
			\begin{figure}[!h]
				\centering
				\includegraphics[width=90 mm]{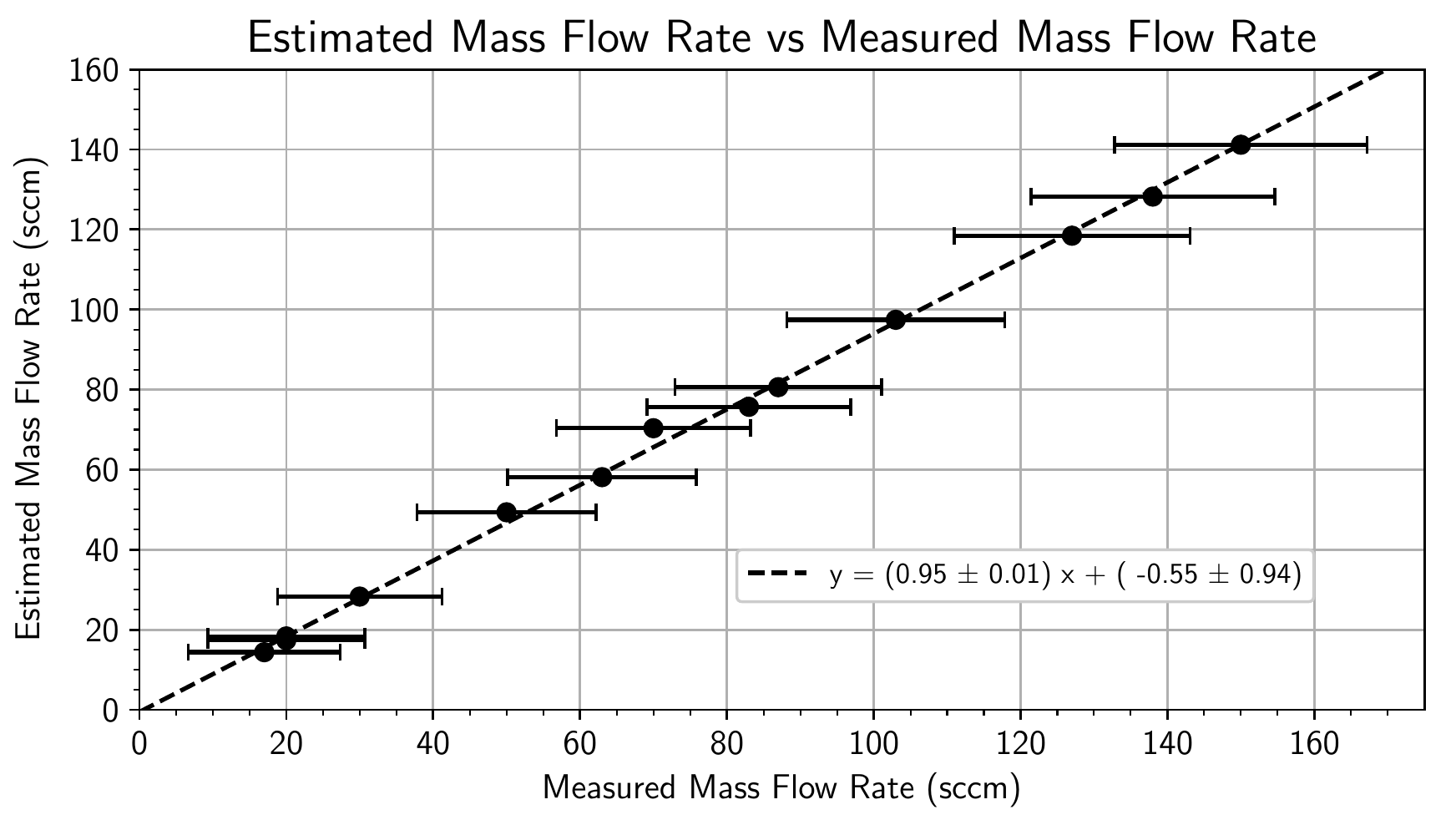}
				\caption{MFC calibration.  Pressure data are used to estimate the actual mass flow rate for a given readout rate.  The different symbols represent different power supplies for the same mass flow controller.  The fitted results show a small deviation from unity that is well within the expected precision of the device.  This fit includes the systematic uncertainty of the MFC calibration, but does not include uncertainties from the volume calculation and possible variation due to temperature.}
				\label{fig:MFC_Calibration}
			\end{figure}
			
			At each MFC dial setting $\dot{m}_\mathrm{meas.}$ we derive an estimated mass flow $\dot{m}_\mathrm{Est.}$ from the pressure fits using Eqn.~\ref{eq:massfloweqnAtest}.  The results (Fig.~\ref{fig:MFC_Calibration}) are linear within the assumed statistical uncertainties, yielding a best-fit deviation of 5~\% between reported and true mass flows:
			\begin{equation}
				\label{eq:MFC_Calibration}
				\dot{m}_\mathrm{Est}=(0.945\pm 0.013)\dot{m}_\mathrm{meas.}-(0.55\pm 0.94)~\mathrm{sccm}~\mathrm{(stat.)}.
			\end{equation}
			Additional systematic errors come from the volume of the turbopump and temperature of the gas.  The system ought to be in thermal equilibrium with the room, but we note that 5~$^{\circ}\C$ change in room temperature would change the resulting slope in Fig.~\ref{fig:MFC_Calibration} by 0.017.  While pressure and timing resolution contribution to the error budget is negligible, the volume estimation provides a significant uncertainty.  The assumed 5~\% uncertainty leads to an additional uncertainty in the slope of 0.018.  These uncertainties are uncorrelated and added in quadrature.  Including the systematic terms, the slope of the linear fit is
			\begin{equation}
				d\dot{m}_{Est}/d\dot{m}_{meas.}=0.945\pm 0.013~(\mathrm{stat.})\pm 0.024~(\mathrm{sys.}),
			\end{equation} which suggests a moderate deviation between true and ``dial" mass flow.  This is within the expected range of 5.32~$\%$ from the MFC manual.  The MFC is confirmed to be calibrated to within the vacuum device's intrinsic precision.
			
			\subsection{Equilibrium Pumping Tests}
			\label{subsec:Test 3: Equilibrium Pumping Tests}
			
			In the remaining test, we verified the pumping speed of the various pumps by flowing gas into the target chamber with a pump running, allowing the system to settle into a dynamic equilibrium.  This most closely mimics the state of the chamber during data-taking, and the pressure profiles achieved will keenly depend on the effective pumping speeds of the pumps in the assembled system.  We selectively operated the Roots pumps one at a time by opening their respective valves, and for each configuration recorded the equilibrium pressure profiles for 10 mass flow rates from 100 to 1000~sccm.  
			
			The pumping speed, in volume per unit time, can be related to mass flow and gas density as follows, noting the relationship between pressure and mass density:
			\begin{align}
				\dot{m}=&\rho S\\
				P =& \rho R_s T  ,
			\end{align}
			where $R_s=R/M$ is a specific gas constant of hydrogen with the gas constant $R=8.314~\J/\mol/\K$ and the molar mass $M =2.02\times 10^{-3}$~kg/mol.  In dynamic equilibrium, an arbitrary volume has conductance $C$, defined as the relation between $dN/dt$, the number flux of molecules in through the inlet of the volume (which is identical to the flux through the outlet), and $n_{1,2}$, the number densities at the inlet and outlet:
			\begin{equation}
				\frac{dN}{dt}=C(n_1-n_2)
			\end{equation}
			\begin{align}
				\label{eq:numberandpumpingspeed}
				\frac{dN}{dt}=&S_1 n_1= S_2 n_2   .
			\end{align} 
			By definition, the pumping speed $S$ is identical with volume flow rate with subscripts 1 for inlet and 2 for outlet.  Combined, the equations lead to the relationship between volume flow and conductance:
			\begin{equation}
				\label{eq:effectivepumpingspeed}
				\frac{1}{S_1}=\frac{1}{S_2}+\frac{1}{C} .
			\end{equation}
			In order to determine the effective pumping speed $S_1$ of the inlet, we treat the outlet as the pump location with known pumping speed $S_2=S_p$ and the inlet as the pressure gauge's location with known pressure:
			\begin{align}
				\label{eq:densityandmassflowrate}
				\rho_1=&M n_1\\
				\label{eq:pumpingspeedandmassflowrate}
				S_1 =& \dot{m}/\rho_1 =\dot{m} R_s T /P_1  .  
			\end{align}
			The conductance of an arbitrary pipe can be expressed as a function of the physical geometry $G$,
			characterized by parameters such as radii and length, and pressure conditions $P_1$ and $P_2$, written $C(G,~P_1,~P_2)$.  Analytic derivations are not always possible, but heuristic equations exist for many simple cases \cite{Roth}.  Combining Eqns.~\ref{eq:effectivepumpingspeed} and
			\ref{eq:pumpingspeedandmassflowrate} with this in mind, we can write:
			\begin{align}
				\label{eqn:Pumping_Speed_and_Pressure}
				\frac{\dot{m}R_s T }{P_1}=\frac{1}{S_p}+\frac{1}{C(G,~P_1,~P_2)}.
			\end{align}
			In a steady-state system with one inlet and one outlet, the mass flow into the system must be the same as the mass flow through the pump.  Using this, we can determine the pressure at the pump from the published pump speed curve (see \ref{subsec:Pumping Speed of the Roots Blower iH1000}).  This gives two ways of computing the conductance:  the effective conductance of the volume between the gauge and the pump can be determined via either heuristics (assuming the geometry is simple) or simulation, and compared to the conductance inferred from the measured pressures, flows, and pump speed curves.
			
			We note that both heuristics and simulation require knowledge of the mechanics of gas flow in the region, characterized by the Knudsen number, $\mathrm{Kn}=\lambda/L$, which describes the collective nature of the flow, and the Reynolds number, $\mathrm{Re}=\rho u L/ \mu$, which describes the degree of turbulence in the flow.  For the target chamber, the pertinent length scale is set by the apertures in the baffles, $L\sim 3~ \mm$. The remaining terms, mean free path $\lambda$, density $\rho$, and drift velocity $u$ are intrinsic features of the gas: $\lambda \sim 10^{-4}/(P~[\mbar])$,  $\rho \sim 10^{-4}\times (P~[\mbar])~\kg/\m^3 $, $ u \sim 300~\m/\s$, $\mu \sim 9 \times 10^{-6}~\kg/\m/\s$ at room temperature, and so Kn$\sim 0.04/(P~[\mbar])$ and Re $\sim 10\times P~[\mbar]<1000$.  The pertinent gas flow regime is laminar, with continuous flow in the central part of the target, giving way to intermediate flow in the baffle regions and molecular in the vicinity of the turbopumps and beyond, as expected.
			
			\begin{figure}[!h]
				\centering
				\includegraphics[width=90 mm]{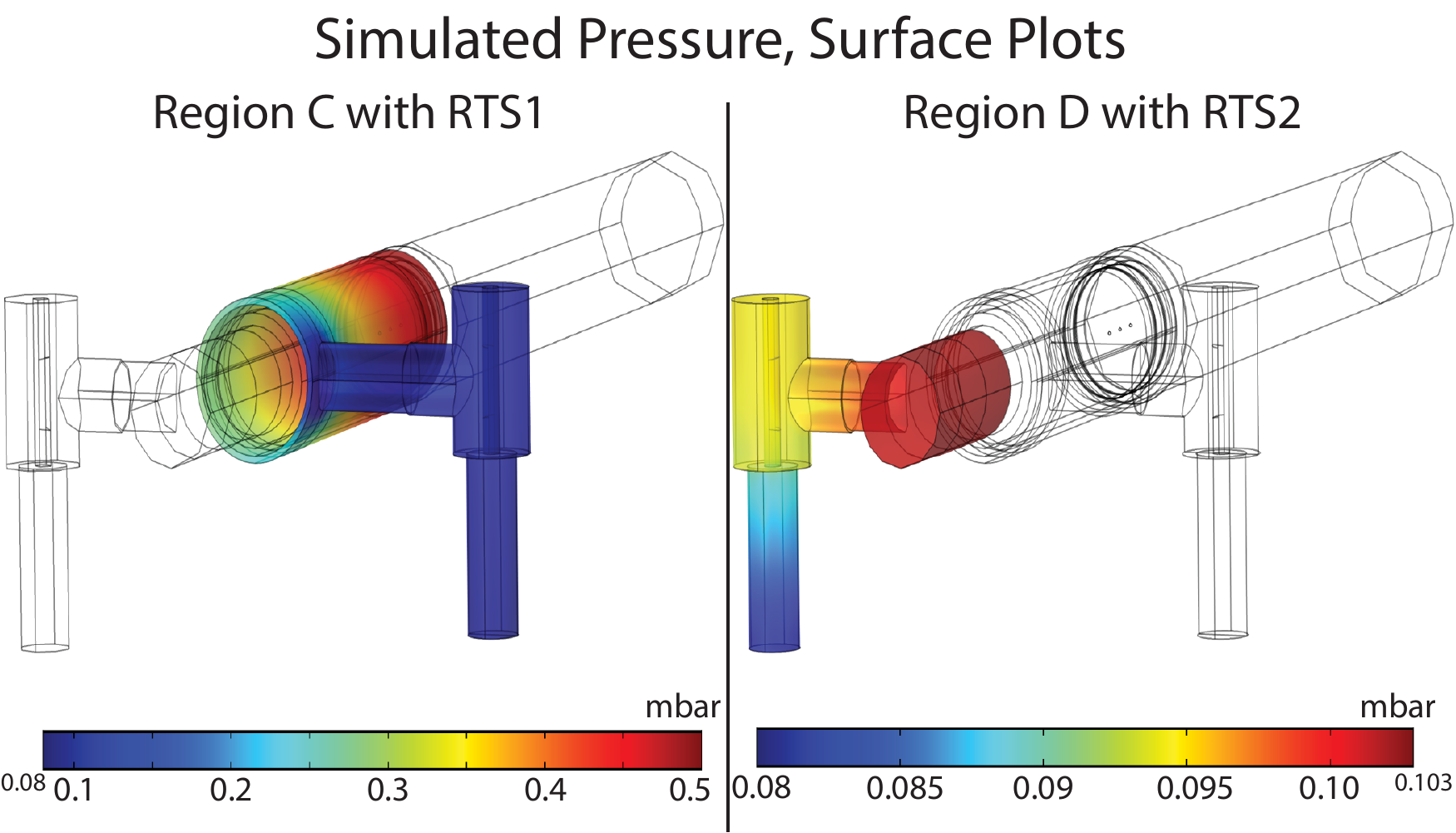}
				\caption{Snapshots of COMSOL simulation results where $\dot{m}=$ 1000 sccm.  Presented are the pressure profiles in the volumes leading to the first (left) and second (right) Roots pumps.  The annular region (left) shows how asymmetric geometries can have large variations in pressure along the azimuth.}
				\label{fig:COMSOL}
			\end{figure}
			
			Finite element analysis (FEA) simulation provides another means of determining $P_1$ and $P_2$ from $\dot{m}$ and $S_p$, particularly useful for more complex geometries where heuristics do not exist.  
			The simulation takes input parameters of the mass flow rates at the inlet and the pressure at the outlet.  For a fixed mass flow rate, several outlet pressure setups were tested.  This allows us to find the configuration where the pressure at the outlet, multiplied by the nominal pumping speed at that pressure, matches the input mass flow rate.  From this configuration, we can extract the pressure at the location of the gauges and compare these values to the measurements in the physical setup.
			The simulated pressures along the surface of some portions of the vacuum system are shown in Fig.~\ref{fig:COMSOL}, which highlight the pressure differentials in azimuthally asymmetric volumes that are not captured with heuristic models.
			\begin{figure}[!h]
				\centering
				\includegraphics[width=90 mm]{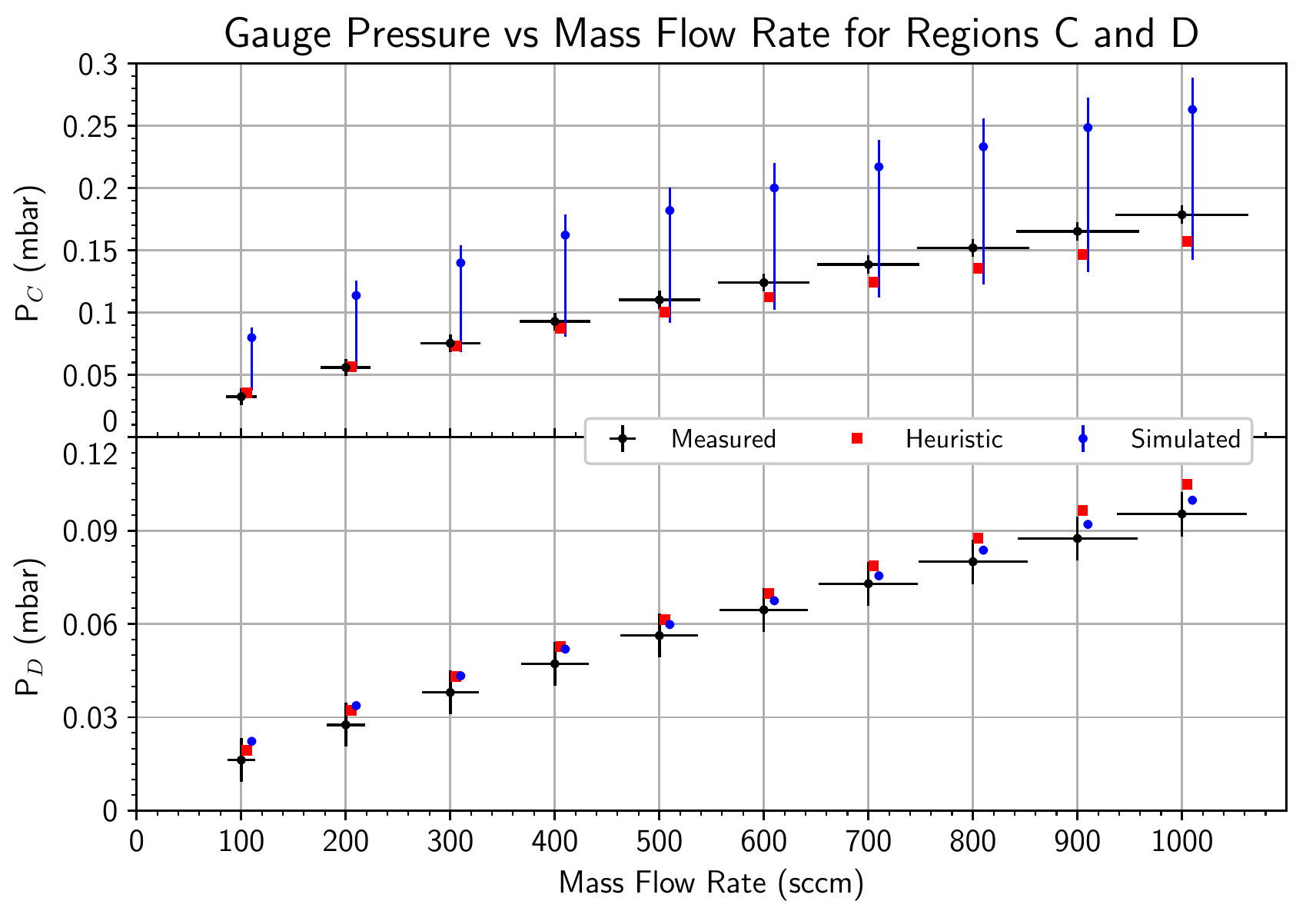}
				\caption{
					Gauge pressure vs mass flow rate for regions C and D, showing measurements, heuristic calculation, and COMSOL simulation.  For clarity, the mass flow rates for analytic calculation and simulation are shifted by 10~sccm and 20~sccm, respectively.  Measured pressure data points are shown in black, with error bars.  Analytic calculations, using heuristics to estimate the pressure drop between gauge and pump, are marked with red squares.  Simulation results are marked with blue circles with vertical error bars.
				}
				\label{fig:Pressure_Simulation}
			\end{figure}
			
			Measured, analytically corrected, and simulated results can be found in Fig.~\ref{fig:Pressure_Simulation}.  The pressure in the annular region C2 changes significantly between the location of the gauge, the inlets, and the pipe leading to the pump.  The error bars shown for that plot mark the highest (farthest from the valve) and lowest pressures (at the valve) along the circumference of the region, with the data point marking the pressure at the gauge location.  Though the conductance of an annular pipe can be calculated heuristically, the spread of pressures due to asymmetries along the azimuth of the pipe are not represented.  In contrast, simulations show that the pressure does not vary sharply in the region D2, which follows expectations of the simpler, large-conductance geometry.  In both cases, the COMSOL simulated and analytically corrected pressure data show a good match with the measured pressure.  

			\section{Conclusion}
			In this article, we presented a technical description of the DarkLight windowless hydrogen gas target.  In addition, the commissioning operation of the target at Jefferson Lab's LERF was described.  Based on the results of that run, several improvements were identified, many of which were implemented in the part of the gas system which was assembled and operated at Bates.  
			We further verified the calibration of all gas system components with minimal assumptions, and compared heuristic and FEA models of gas pressure profiles in support of that calibration.
			
			The data from the Bates and LERF tests demonstrate that a windowless gas target can be constructed of sufficiently high density to meet the requirements of the DarkLight experiment.  This 
			work provides confidence that the targets required for future experiments at ERLs currently under construction \cite{Hoffstaetter:2018ila, Hug:2017ypc} are technically feasible.
			
			\section{Acknowledgements}
			\label{sec:Acknowledgement}
			This research is supported by the NSF
			MRI Program (Award No.~1437402), the U.S. Department of
			Energy Office of Nuclear Physics (Grant No.  DE-FG02-
			94ER40818 and DE-AC05-06OR23177), and the U.S.  Department of Energy Office of High
			Energy Physics (Grant No.DE-SC0011970). 
			\appendix
			\section{Appendix}
			\label{sec:Appendix}
			\subsection{Uncertainty of Pressure Measurements}
			\label{subsec:Uncertainty of Pressure Measurements}
			\begin{table}[ht!]
				\centering
				\caption{Error Budget of Vacuum Gauge Devices.  
				}
				\begin{tabular}{|c|c|c|c|}
					\hline
					Device & Type & in Eqn.~\ref{eq:pressureerroreqn}& Effects \\
					\hline
					\multirow{4}{*}{CDG}& Accuracy&$\sigma_{\CDG.~A}$ & 0.2~$\%$\\
					&Offset &$\sigma_{\CDG.~O}$ & 6.7$\times$10$^{-3}$~mbar\\
					&Gain &$\sigma_{\CDG.~G}$ & 0.1~$\%$\\ 
					&Resolution&$\sigma_{\CDG.~R}$ & 4$\times$10$^{-4}$~mbar \\
					\hline
					\multirow{3}{*}{VGC}& Gain&$\sigma_{\VGC.~G}$ & 1.3$\times$10$^{-3}$~mbar\\
					&Offset&$\sigma_{\VGC.~O}$& 1.3$\times$10$^{-3}$~mbar\\
					&Resolution&$\sigma_{\VGC.~R}$ & 1.3$\times$10$^{-4}$~mbar.\\
					\hline
				\end{tabular}
				\label{tab:pressureerror}
			\end{table}
			
			Here we enumerate the contributions to the intrinsic uncertainties of the two types of gauges used in the DarkLight target system.  Both the CDG and the VGC have intrinsic accuracies and resolution values, which are listed in Table \ref{tab:pressureerror} \cite{CDG025D, VGC503}.  The CDG offset and gain arise from temperature effects.  Finally, the pressure measurement error from different sources can be put together:
			\begin{align}
				\label{eq:pressureerroreqn}
				\sigma_{Gain}^{\CDG}&=\sqrt{\sigma_{\CDG.~A}^2+\sigma_{\CDG.~G}^2}=0.22~\%\\
				\sigma_{Offset}^{\CDG}&=\sqrt{\sigma_{\CDG.~O}^2+\sigma_{\CDG.~R}^2+\sigma_{\VGC.~G}^2+\sigma_{\VGC.~O}^2+\sigma_{\VGC.~R}^2} \nonumber\\&=6.94\times10^{-3}~\mbar\\
				P_{true}^{\CDG}&=P_{meas.}^{\CDG}\pm(\sigma^{\CDG}_{Gain}\oplus\sigma^{\CDG}_{Offset})\nonumber\\
				&=P_{meas.}^{\CDG}\pm (0.22~\% \oplus 6.94\times10^{-3}~\mbar).
			\end{align}
			
			The BPG has relatively simple error budget: 15$\%$ of the measurement, so,
			\begin{align}
				\label{eq:pressureerroreqn2}
				P_{true}^{BPG}&=P_{meas.}^{BPG} \pm (0.15 \% \oplus \sqrt{\sigma_{\VGC.~G}^2+\sigma_{\VGC.~O}^2+\sigma_{\VGC.~R}^2}) \nonumber\\
				&=P_{meas.}^{BPG}\pm (0.15 ~\%\oplus1.89\times10^{-3}~\mbar).
			\end{align}
			
			\subsection{Uncertainty of the Mass Flow Rate}
			\label{subsec:Uncertainty of MFC}
			\begin{table}[ht!]
				\centering
				\caption{Error Budget of MFC.}
				\begin{tabular}{|c|c|c|c|}
					\hline
					Device &Type & in Eqn.~\ref{eq:GCF}& Effects \\
					\hline
					\multirow{6}{*}{MFC}& \multirow{3}{*}{Accuracy}&\multirow{3}{*}{$\sigma_{\MFC.~A}$} & 0.3~$\%$\footnote{corresponding readout range: 0--20~sccm}\\
					&&&2~sccm\footnote{corresponding readout range: 20--200~sccm}\\
					&&&10~sccm\footnote{corresponding readout range: 200--1000~sccm}\\
					&Offset&$\sigma_{\MFC.~O}$ & 5~sccm \\
					&Gain&$\sigma_{\MFC.~G}$ & 8~sccm\\        
					&Resolution&$\sigma_{\MFC.~R}$ & 1~sccm\\
					\hline
					&Display Accuracy&$\sigma_{\PS.~D}$ &0.1~$\%$\\
					Power&Accuracy&$\sigma_{\PS.~A}$&5~$\%$\\
					Supply&Resolution&$\sigma_{\PS.~R}$ & 0.5~sccm\\
					&Gas Correction Factor &$\sigma_{\PS.~G}$ & 1.8~$\%$\\
					\hline
				\end{tabular}
				\label{tab:MFC Error}
			\end{table}
			
			The MFC manual indicates that a nonzero gauge scale factor results in a 5~$\%$ uncertainty in the MFC readout \cite{246C,GE50A}.  The scale factor is the product of the gauge factor and the gas correction factor and is an input to the MFC via a control knob.  The gauge factor only depends on the MFC type, and is exactly 1.0 for 1000~sccm full scale, which is used for the tests.  The gas correction factor (GCF) depends on the gas type and the temperature, and is given as follows for hydrogen (see Table \ref{tab:MFC Error}).
			\begin{equation}
				\label{eq:GCF}
				\mathrm{GCF}=1.011\times \frac{T (\K)}{273.15 \K}.
			\end{equation}
			This leads to a 1.8~$\%$ uncertainty from a 5~${}^{\circ}$C uncertainty in temperature.  As in the CDG case, the MFC offset and gain arise from temperature effects.  We estimate the uncertainty of the MFC readout as shown in \ref{subsec:Uncertainty of Pressure Measurements}.  The accuracy varies over the mass flow rate in the mid range mass flow rate, 20--200~sccm as follows.
			\begin{align}
				\label{eq:MFC Readout Uncertainty}
				\sigma_{Gain}&=\sqrt{\sigma_{\PS.~D}^2+\sigma_{\PS.~A}^2+\sigma_{\PS.~G}}=5.32~\%\\
				\sigma_{Offset}&=\sqrt{\sigma_{\MFC.~A}^2+\sigma_{\MFC.O}^2+\sigma_{\MFC.G}^2+\sigma_{\MFC.R}^2+\sigma_{\PS.~R}^2}\nonumber\\&=9.7~\sccm.
			\end{align}
			In all ranges,
			\begin{align}
				\label{eq:MFCerroreqn}
				\dot{m}_{true}&=\dot{m}_{meas.}\pm(\sigma^{\MFC}_{Gain}\oplus\sigma^{\MFC}_{Offset})\nonumber\\   
				&=
				\begin{cases}
					\dot{m}_{meas.}\pm (5.33~\% \oplus 9.5 ~\sccm)        &(\dot{m}_{meas.}<20) \\
					\dot{m}_{meas.}\pm (5.32~\% \oplus 9.7 ~\sccm)           &(20\leq \dot{m}_{meas.}<200)\\
					\dot{m}_{meas.} \pm (5.32~\% \oplus 13.8 ~\sccm)  &(\dot{m}_{meas.}\geq200).
				\end{cases}
			\end{align}
			
			\subsection{Pumping Speed of the Edwards iH1000}
			\label{subsec:Pumping Speed of the Roots Blower iH1000}
			\begin{figure}[!h]
				\centering
				\includegraphics[width=90 mm]{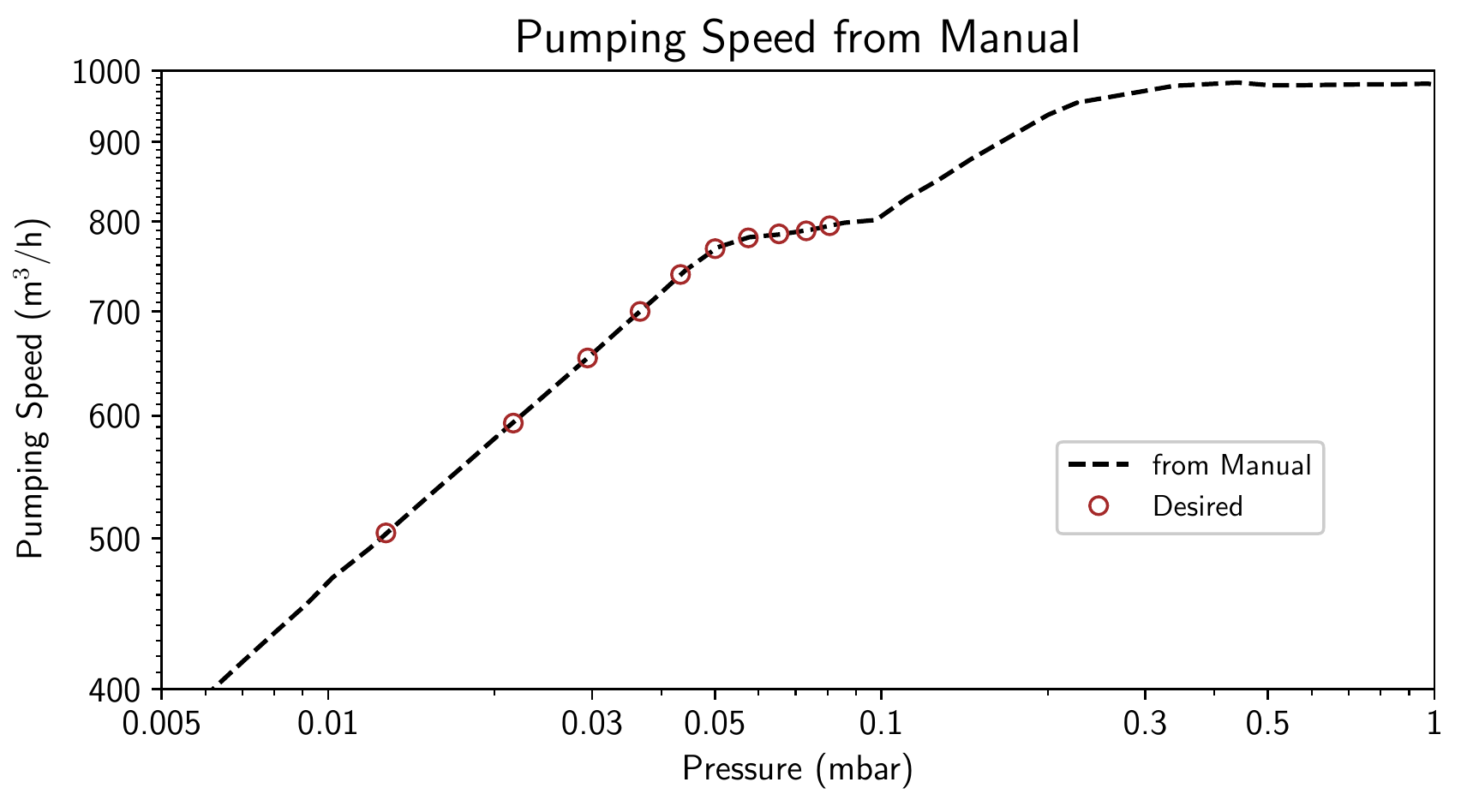}
				\caption{A plot of the pumping speed w.r.t.  the operational pressure of the RTS is presented.  The black dotted curve comes from the manufacturer's guide.  The brown open circles on the curve stand for the desired pressure and pumping speed of the selected mass flow rates.}
				\label{fig:PumpingSpeed}
			\end{figure}
			
			The Roots blowers have pumping speed, $S$, dependent on the pressure, $P_2$, at their inlet.  Since Eqn.~\ref{eqn:Pumping_Speed_and_Pressure}  requires one of these to be known, it is difficult to compare the measured and expected pumping speeds.  The nominal curve relating $S$ and $P_2$ (shown in Fig.~\ref{fig:PumpingSpeed}) could be parameterized, but in practice it is simpler to mark the working points of the pressure tests on this curve instead.  The brown circles, from left to right, represent mass flows through the pump of 100--1000~sccm in increments of 100~sccm.  If the physical pump matches the manual's stated performance, the pressure at the pump inlet, determined from the calibration procedure, must agree with the marked points.
			\bibliographystyle{elsarticle-num}

			
			
			
		\end{document}